\documentclass[conference]{IEEEtran}

\usepackage[final]{graphicx}
\usepackage{amsmath,amsfonts,amssymb}
\usepackage{citesort}

\newcommand{\figurepath}{pix}

\newcommand{\papertitle}{A Tree Search Method for Iterative Decoding
  of Underdetermined Multiuser Systems}

\newcommand{\reals}{\mathbb{R}}
\newcommand{\prob}{\mathbb{P}}
\newcommand{\vect}[1]{\mathbf{#1}}

\newcommand{\expect}[1]{\mathsf{E}\left[#1\right]}

\newcommand{\card}[1]{\left|#1\right|}
\newcommand{\set}[1]{\mathcal{#1}}

\renewcommand{\u}{\vect{u}}
\newcommand{\y}{\vect{y}}
\renewcommand{\r}{\vect{r}}

\newcommand{\z}{\vect{z}}

\renewcommand{\d}{\vect{d}}

\newcommand{\dpr}{\vect{d}^\prime}
\newcommand{\G}{\vect{G}}

\renewcommand{\P}{\set{P}}
\newcommand{\Pmax}{\set{P}_{\textrm{max}}}
\newcommand{\Pmin}{\set{P}_{\textrm{min}}}

\newcommand{\norm}[1]{\left\|#1\right\|}

\newcommand{\eye}{\vect{I}}

\newcommand{\complex}{\mathbb{C}}
\newcommand{\s}{\vect{s}}

\renewcommand{\c}{\vect{c}}
\newcommand{\M}{\set{M}}

\newcommand{\D}{\set{D}}

\newcommand{\C}{\set{C}}

\newcommand{\T}{\vect{T}}

\newcommand{\Real}{\Re e}

\renewcommand{\S}{{\mathcal{S}}}

\newcommand{\diag}{{\mathrm{diag}}}
\newcommand{\one}{\vect{1}}

\renewcommand{\S}{{\vect{S}}}

\title{\papertitle}
\author{
\authorblockN{Adriel P. Kind} 
\authorblockA{Agere Systems Australia}
\and
\authorblockN{Alex Grant}
\authorblockA{Institute for Telecommunications Research\\
  University of South Australia}
}
 
\begin{document}
\maketitle
\def\thefootnote{}
\footnotetext{This work was undertaken while A. Kind was at the Institute for 
    Telecommunications Research.  A. Grant is supported in part by the
  Australian Government under grant DP0344856.}
\begin{abstract}
  Application of the turbo principle to multiuser decoding results in
  an exchange of probability distributions between two sets of
  constraints. Firstly, constraints imposed by the multiple-access
  channel, and secondly, individual constraints imposed by each users'
  error control code. A-posteriori probability computation for the
  first set of constraints is prohibitively complex for all but a
  small number of users.  Several lower complexity approaches have
  been proposed in the literature. One class of methods is based on
  linear filtering (e.g. LMMSE).  A more recent approach is to compute
  approximations to the posterior probabilities by marginalising over
  a subset of sequences (list detection). Most of the list detection
  methods are restricted to non-singular systems. In this paper, we
  introduce a transformation that permits application of standard
  tree-search methods to underdetermined systems. We find that the
  resulting tree-search based receiver outperforms existing methods.
\end{abstract}

\section{Introduction}\label{sec:intro}

It is well known that joint decoding can improve performance in
multiple-access systems.  Joint maximum likelihood (ML) decoding,
which minimizes the overall probability of error is however
prohibitively complex~\cite{GiaWil96TCa}. Brute force computation of
the jointly ML codeword sequences for $K$ users is $O(Q^{K\kappa})$
for $Q$-ary modulation and constraint length $\kappa$ codes.

The good performance and low complexity of the turbo decoder
\cite{BerGlaThi93} led to application of the turbo principle to joint
multiuser decoding.  Figure~\ref{fig:decoder} shows a schematic
representation of the ``canonical'' iterative multiuser
decoder~\cite{ReeSch98TC,ValWoe98PIMRC,Moh98TC,MohGul98IT}.  This
decoder treats the users' forward error correction codes as an ``outer
code'' and the interdependency introduced by the multiple access
channel as an ``inner code''. The decoder iterates between
a-posteriori probability (APP) computation for the inner code and
individual APP decoding of each user's FEC code. The multiuser APP
computation is $O(Q^{K})$, an improvement over joint ML decoding, but
still prohibitive.

One low complexity alternative is to replace the inner APP decoder
with a linear filter. Examples include soft interference
cancellation~\cite{Hag96,AleGraRee98} and linear minimum mean-squared
error filtering~\cite{WanPoo99}.  These approaches can work quite
well, but there is still room for improvement compared to the exact
computation of the multiuser APP.

A more powerful approach is to compute an approximation of the
multiuser APP by marginalizing over a subset of sequences (in many
cases only a small subset is required), found using various different
list
detectors~\cite{ReiGraAle02,ReiGraKin03,Hag03,baro,lsd,BouGre03,KuhHag04,sdprior}.
Most of these list-detection based methods rely on Cholesky
decomposition of the correlation matrix as a first step. In
underdetermined systems (more users than signaling dimensions), this
decomposition cannot be performed. First steps towards avoiding this
problem have been made in~\cite{DamElGCai04ciss,DamElGCai04isit},
based on filtering, followed by lattice reduction.

The main contribution of this paper is a simple transformation which
creates a virtual full rank system, which permits Cholesky
decomposition and the straight-forward application of tree-search
methods in underdetermined multiuser systems. The method requires less
computational complexity that the one described
in~\cite{DamElGCai04ciss,DamElGCai04isit}.  Numerical results
demonstrate the superior performance of the approach, which is
compared to other techniques from the literature.

\section{System Model and Canonical Decoder\label{sec:sysmodel}}

Consider a multiple-access system with $K$ radio terminals, or users,
simultaneously transmitting forward error correction coded digital
data across an additive white Gaussian noise (AWGN) channel.
The encoder for user $k=1,2,\dots,K$ operates as follows.  A length
$I$ frame of independent equi-probable information bits $\u_k$ is
encoded by a rate $R_\C$ code $\C_k$.  The $I/R_\C$ coded bits $\c_k$
are then permuted with the interleaver $\Pi_k$, and parsed into length
$\log Q$ segments.  These segments are mapped onto a stream $\d_k$ of
$I/R_\C\log Q$ constellation symbols according to some memoryless
mapping, and then multiplexed onto the symbol sequences of length
$I/R_\C\log Q$.  Each user transmits at a rate of $R_\C\log Q$ bits
per channel use.

A data vector $\d=(d_{1},\cdots,d_{K})^T\in \D^{K}$ represents all
users' symbols in a given symbol interval (assuming symbol synchronous
transmission for simplicity of explanation).  The complex
constellation $\D\subset\complex$ has $\card{\D}=Q$ unique elements,
with moment constraints $\expect{\d}=\vect{0}$, and
$\expect{\d\d^*}=P\eye_K$, and symbols are equiprobable.  The average
transmit power per user is $P$.

Each symbol is multiplied by a length $L$ modulation vector $\s_k$,
which has real random elements chosen uniformly from $\pm 1/\sqrt{L}$.
A vector $\z\in\complex^L$ with independent white zero-mean Gaussian
element represents thermal noise with variance $\sigma^2$ per real
dimension.  In a coded system where each user employs a rate $R$ code
and transmits with power $P$, the appropriate signal-to-noise measure
we will use is $E_b/N_0=P/2\sigma^2R\log Q$.

We assume that each user's signals are received with
identical power, phase and delay, although these are not fundamental
restrictions imposed by the proposed receiver.  After standard
manipulations the multiple-access channel may be represented by
\begin{equation}\label{eq:sysmodel}
  \r=\S\d+\z
\end{equation}
where $\S=(\s_1,\cdots,\s_K)\in\{\pm 1/\sqrt{L}\}^{L\times K}$.
We only consider the case that $K>L$, ie. the number 
of users exceeds the number of independent observations.

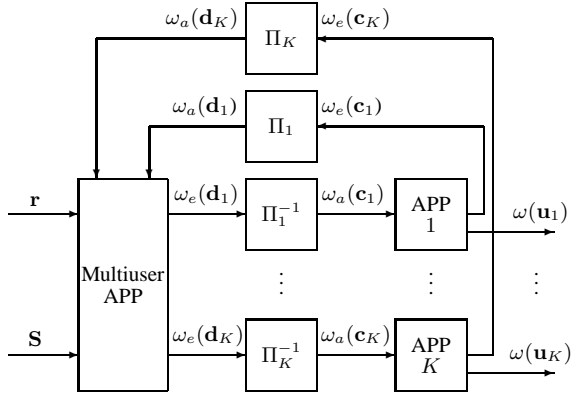
\begin{figure}[htbp]
  \centering\small\setlength{\unitlength}{1.3cm}
  \scalebox{0.9}{
  \begin{picture}(6.2,4.8)(.375,-.15)
    \put(.2,.4){\vector(1,0){.8}}
    \put(.2,2){\vector(1,0){.8}}
    \put(1,0){\framebox(1,2.4){\shortstack{Multiuser\\APP}}}
    \put(2,.4){\vector(1,0){.9}}
    \put(2,2){\vector(1,0){.9}}
    \put(3.7,.4){\vector(1,0){.9}}
    \put(3.7,2){\vector(1,0){.9}}
    \put(2.9,1.6){\framebox(.8,.8){$\Pi^{-1}_1$}}
    \put(2.9,0){\framebox(.8,.8){$\Pi^{-1}_K$}}
    \put(3.3,1.3){\makebox(0,0)[c]{$\vdots$}}
    \put(4.6,1.6){\framebox(.8,.8){\shortstack{APP\\$1$}}}
    \put(4.6,0){\framebox(.8,.8){\shortstack{APP\\$K$}}}
    \put(5,1.3){\makebox(0,0)[c]{$\vdots$}}
    \put(5.4,.2){\vector(1,0){1}}
    \put(5.4,1.8){\vector(1,0){1}}
    \put(5.4,2){\line(1,0){.2}}
    \put(5.6,2){\line(0,1){1}}
    \put(5.6,3){\vector(-1,0){1.9}}
    \put(2.9,3){\line(-1,0){1.1}}
    \put(1.8,3){\vector(0,-1){.6}}
    \put(5.4,.4){\line(1,0){.3}}
    \put(5.7,.4){\line(0,1){3.6}}
    \put(5.7,4){\vector(-1,0){2}}
    \put(2.9,4){\line(-1,0){1.7}}
    \put(1.2,4){\vector(0,-1){1.6}}
    \put(2.9,2.6){\framebox(.8,.8){$\Pi_1$}}
    \put(2.9,3.6){\framebox(.8,.8){$\Pi_K$}}
    \put(.5,.5){\makebox(0,0)[b]{$\S$}}
    \put(.5,2.1){\makebox(0,0)[b]{$\r$}}
    \put(2.47,.5){\makebox(0,0)[b]{$\omega_e(\d_K)$}}
    \put(2.45,2.1){\makebox(0,0)[b]{$\omega_e(\d_1)$}}
    \put(4.15,.5){\makebox(0,0)[b]{$\omega_a(\c_K)$}}
    \put(4.12,2.1){\makebox(0,0)[b]{$\omega_a(\c_1)$}}
    \put(6.25,.3){\makebox(0,0)[b]{$\omega(\u_K)$}}
    \put(6.25,1.9){\makebox(0,0)[b]{$\omega(\u_1)$}}
    \put(4.15,4.1){\makebox(0,0)[b]{$\omega_e(\c_K)$}}
    \put(4.1,3.1){\makebox(0,0)[b]{$\omega_e(\c_1$)}}
    \put(2.4,4.1){\makebox(0,0)[b]{$\omega_a(\d_K)$}}
    \put(2.43,3.1){\makebox(0,0)[b]{$\omega_a(\d_1$)}}
    \put(6.2,1.3){\makebox(0,0)[c]{$\vdots$}}
  \end{picture}}
  \caption{Iterative joint-APP multi-user receiver.}
  \label{fig:decoder}
\end{figure}

The canonical iterative decoder is shown in Fig.  \ref{fig:decoder}.
The goal is to infer the value of $\u_k$, $k=1,\cdots,K$, based on
$\r$, $\S$ and the constraints $\C_k$. The module labelled Multiuser
APP computes the marginal posterior probability matrix
$\omega(\d)\in\prob^{Q\times K}$, which has as columns the probability
mass functions for the corresponding symbols, based on all the
available information using the constraint \eqref{eq:sysmodel}, as
well as the prior probability matrices $\omega_a(\d)$. This inner
decoder is the focus of this work.

The other constraint, separated from the first by an interleaver, is
the single-user decoders, which calculate extrinsic probabilities
$\omega_e(\c_k)$ based on the codes and $\omega_a(\c_k)$ for all $k$.  
The process
repeats by iteratively exchanging information in the form of extrinsic
probability matrices between the two modules.  The individual APP
decoders also compute, on the final iteration, the data sequence
probabilities $\omega(\u_k)$.  

The computation of the marginal symbol posteriors in the APP detector
entails marginalising the joint probability $p(\d,\r|\r,\S,N_0)$ over
\emph{each} possible sequence $\d$. Brute force marginalisation for
each user is therefore a summation with $Q^{K-1}$ terms, clearly
impractical for all but small numbers of users. In practice however,
nearly all of the probability for systems of interest is contained in
a relatively small subset of $\P$ of those terms \cite{KinGra04isit},
and provided those terms can be isolated, the marginalisation can
become computationally tractable. This is the idea goes back
to~\cite{WeiRasWyr97} (in an uncoded context) and has seen a recent
revival in the framework of iterative
processing~\cite{ReiGraAle02,ReiGraKin03,Hag03,baro,lsd,BouGre03,KuhHag04,sdprior}.
The goal of the next section is to approximate this sum with greatly
reduced complexity.

\section{Approximation of the Multiuser APP}\label{sec:app}

The posterior log joint-probability of a particular hypothesis sequence 
$\dpr$ is equal to
\begin{equation}\label{eq:jointprob}
  \log p(\r,\dpr|\S,N_0)=c-\frac{1}{N_0}\norm{\r-\S\dpr}^2
  +\log p(\dpr)
\end{equation}
where $c$ is a constant and $p(\dpr)$ is the prior probability of the
sequence.  Expand the squared-distance term as
\begin{equation}\label{eq:form1} 
  \begin{split}
    \norm{\r-\S\dpr}^2&=\r^*\r-2\Real\{\r^*\S\dpr\}+{\dpr}^*\norm{\S\dpr}^2\\
    &=c+\Real\{\y\dpr\}+\norm{\S\dpr}^2
  \end{split}
\end{equation}
where $\y=-2\r^*\S$.  In order to simplify the search for sequences
$\dpr$ that minimise \eqref{eq:form1}, a recursive expression in
$d_1^\prime,\cdots,d_K^\prime$ may be obtained if $\G=\S^*\S$ is
positive-definite.  This cannot be the case when $K>L$. In our model
$\S$ is not even guaranteed to have rank $L$.  In order to obtain an
equivalent full-rank system, we exploit the following representation.
\begin{equation*}
  \begin{split}
    &\norm{\S\d}^2=\d^*\G\d=\sum_{k=1}^Kg_{kk}|d_k|^2+d_k^*
    \sum_{\substack{j=1\\j\ne k}}^Kg_{jk}d_j\\
    &=\sum_{k=1}^K\left(\rho_k|d_k|^2+d_k^*\sum_{\substack{j=1\\j\ne k}}^K
    g_{jk}d_j\right)+(g_{kk}-\rho_k)|d_k|^2
  \end{split}
\end{equation*}
where $\rho_k\in\reals$ is a free parameter.  The terms in brackets
define the columns of a new matrix, and a sufficient condition for
that matrix to be positive-definite is that each term is positive
irrespective of $\d$.  The following procedure is used to transform
the log-likelihood into an additive recursive metric with $K$ terms.
This technique, which we believe to be new, is the main contribution
of this paper.
\begin{enumerate}
\item Choose a positive constant $\rho\in\reals^+$ satisfying
  \begin{equation}\label{eq:rho}
    \rho>(K-1)\max_{i,j}
    \frac{-\Real\{\D_i^*\D_j\}}{|\D_i|^2}
  \end{equation}
  where $\D_1,\cdots,\D_Q$ are the elements of $\D$.
\item Construct the vector $\u=\diag(\G)-\rho\one\in\reals^K$,
  where $\one$ is the all-ones vector.
\item Construct a new matrix $\tilde{\G}$ by setting all diagonal elements 
  of $\G$ equal to $\rho$. The matrix $\tilde{\G}$ is guaranteed to be 
  positive-definite.
\item Compute the factorisation $\T^*\T=\tilde{\G}$,
  where $\vect{T}\in\reals^{K\times K}$ is lower-triangular.
\end{enumerate}
By setting the prior term $\mathcal{L}(\dpr)=-N_0\log p(\dpr)$, and assuming
statistical independence of the prior symbol probabilities due to the 
interleaver, \eqref{eq:jointprob} may be written as
\begin{equation}\label{eq:dunno}
  \begin{split}
    -N_0\log p(\r,\dpr|&\S,N_0)= 
    c+\sum_{k=1}^K \Real\{y_kd^\prime_k\}+\\
    &\left|\sum_{j=1}^kt_{kj}d_j^\prime\right|^2+\mathcal{L}(d_k^\prime)+
    u_k|d_k^\prime|^2 
  \end{split}
\end{equation}
For constant energy symbol constellations such as $Q$-ary PSK, the 
last term in \eqref{eq:dunno} is absorbed into $c$, and we only require
$\rho>(K-1)$.

Quadratic forms such as \eqref{eq:dunno} admit a tree
representation~\cite{WeiRasWyr97}.  The equation represents a $Q$-ary
tree of depth $K$, where each of the $Q^{k}$ nodes at depth $k$
represent a partial sequence with an associated positive path weight,
and the leaf nodes represent sequences $\dpr$ with total path weight
equal to $c-N_0\log p(\dpr,\r|\S,N_0)$.  Hence, the problem reduces to
finding the $\P$ leaf nodes in the tree with minimum weight, which may
be approximated using tree search techniques.

The simple manipulations applied to $\G$ described above artificially 
create a virtual full rank channel from a rank-deficient one, assigning a
positive path weight to every node in the tree and allowing sequential
search to be applied to \eqref{eq:dunno}.  This is not to say that
extra information is obtained about the symbols via the
transformations described; only that the information about the
interfering signals is spread out onto a greater number of effective
observations, so that any sequential search techniques developed for a
full rank channel may also be applied in the overloaded or singular
case.

A transformation for overloaded linear systems was presented in 
\cite{DamElGCai04ciss,DamElGCai04isit}, which similarly creates a virtual 
full-rank system to which the full $Q$-ary tree may be assigned.
The approach is based on a minimum mean square error 
generalised decision feedback equaliser filter, followed by lattice 
reduction, column re-ordering, and then triangular factorisation (if
tree/sphere decoding is used).
These transformations are significantly more complex than our procedure, and 
may be unsuitable for time-varying channels.
The approach also colours the noise, so that the system no longer lends 
itself naturally to the iterative APP framework.


A depth-first tree-search was used in~\cite{Hag03,baro,KuhHag04},
which necessitated special treatment of the prior probability on those
paths that did not reach full depth. We propose a breadth-first search
using the $T$-algorithm.  The $T$-algorithm was used in
\cite{And89,Sim90} for near-optimal hard-decision decoding of channel
codes up to a pre-determined minimum-distance with significant
complexity savings over the Viterbi algorithm.  The related
$\M$-algorithm retains exactly $\M$ paths at each depth, regardless of
the actual weights of each partial path (this approach was used
in~\cite{ReiGraAle02}).  In practice the statistical nature of the
noise and the spreading sequences for each transmission may require a
different number of sequences to approximate the APP. It should also
be noted that any other tree search algorithm could be used, with
slightly varying levels of performance and complexity. The key step
is~\eqref{eq:dunno}, which admits such tree representations for
overloaded systems.

We exploit the heuristic observation that paths with very large
partial path weight are unlikely to be components of low weight paths.
Rather than retaining a fixed number of paths at each depth, the
$T$-algorithm attempts to adapt to the channel conditions by only
retaining paths at each depth with weight not exceeding the best
weight by more than $T$, where $T$ is a parameter of the algorithm.
When the algorithm terminates at the leaves, the best $\P$ sequences
are used in the marginalisation.

At low SNR in the early iterations, or with few receiver observations
compared to the number of transmitters, large numbers of paths will
exist with similar path weight.  Due to complexity constraints in this
scenario the number of retained paths at each depth must be limited to
$\Pmax$, and the algorithm essentially becomes the $\M$-algorithm with
$\M=\Pmax$.  In more favourable circumstances however, very few paths
are required and the $T$-algorithm adapts automatically to take
advantage of the conditions with greatly reduced complexity.

When no prior information is available the $T$-algorithm adapts very
well to the channel, automatically finding a good
performance/complexity trade-off through the parameter $T$.  
As a general rule, the $T$-algorithm only tends to approach the $\Pmax$ 
bound in the early iterations, since the search is greatly facilitated 
by the prior probabilities once they become available. When very strong 
prior information is available however, the only sequences
retained will be those dictated by the priors, since other paths will
be discarded in the early depths.  In this case the detector will
glean little new information, and the information about the symbols
will quickly become correlated over iterations.  Hence, another
parameter $\Pmin$ must also be set, forcing the algorithm to consider
a certain minimum number of sequences at each depth.  The effect is
only significant in highly loaded systems where many iterations are
required for convergence.

The $T$-algorithm finds full-depth paths through the tree, and the
prior probabilities are incorporated in a natural fashion. 
In contrast, the depth-first strategy of~\cite{Hag03,baro} required
special handling of the prior probabilities, while the method 
of~\cite{KuhHag04} required an initial breadth-first search in order to 
exploit the priors during the main search. 
The $\M$-algorithm based approach used in~\cite{ReiGraAle02,ReiGraKin03} 
did not directly incorporate priors (this was done in a separate combining 
step). 
Other approaches incorporate the prior probability into spherical or
branch-and-bound decoders in various ways~\cite{lsd,BouGre03,sdprior}, but 
tend to be quite complex and unsuitable for large-dimensional systems.

The receiver complexity is dominated by Cholesky factorisation of the
$K\times K$ matrix $\tilde{\G}$, and then by the tree search during the 
iterations. 
The complexity of the $T$-algorithm is upper bounded by $K\Pmax$ node
computations per iteration, but this bound only ever tends to be reached
in the early iterations, as discussed above. 
Contrast this for example with the LMMSE filter \cite{WanPoo99} as the 
inner detector, which requires an initial matrix inverse, and then a matrix
inverse \emph{per user} on each subsequent iteration.

\section{Numerical Results}\label{sec:simulations}
In this section we consider a benchmark model, with length $L=8$
random PN spreading sequences and no fading.  The model is difficult
to work with, since a significant probability exists that the
spreading matrix $\S$ will have linearly dependent rows.

The individual users transmit BPSK symbols, which are encoded with a 
nonsystematic 4-state rate 1/2 convolutional code, described by the 
feed forward generator polynomials $(05,07)$.
A length $2I=1000$ interleaver between the encoder and the transmitter is 
generated randomly for each user.

We consider joint iterative decoding of the system using the
$T$-algorithm, where $\Pmax$ is set to $512$ and the threshold $T$ is
set to $16N_0$.  
The bound $\Pmax$ is deliberately set large in order to demonstrate the
performance advantage of closely approximating the APP.
The $T$-algorithm for the overloaded case is
furnished by the matrix manipulations proposed in Section
\ref{sec:app} for computing the log-likelihood.

\begin{figure}[htbp]
  \centering\small\setlength{\unitlength}{.74mm}
  \scalebox{.95}{
  \begin{picture}(100,80)
    \put(0,3){\includegraphics[width=74mm]{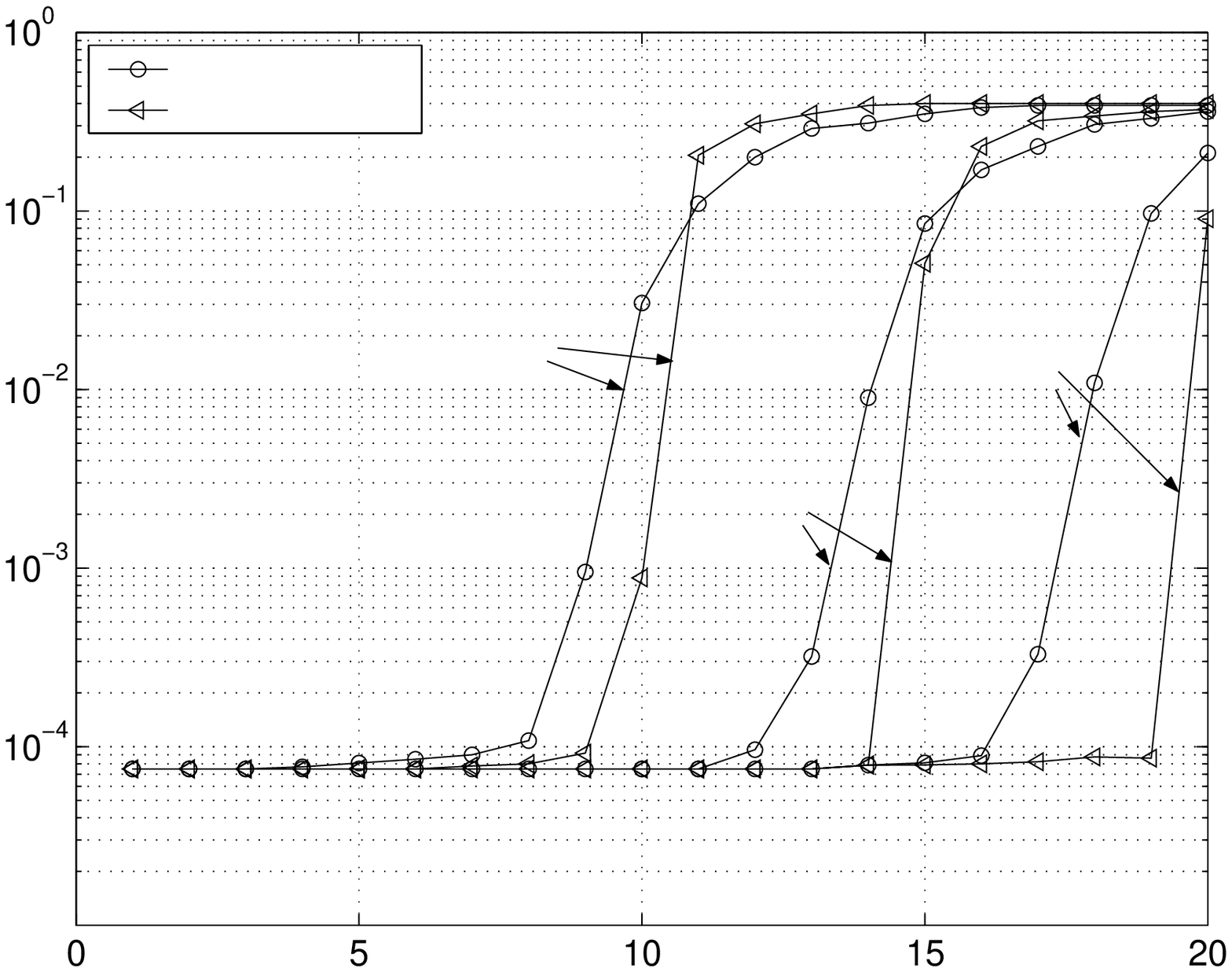}}
    \put(23,77.3){\makebox(0,0)[t]{\scriptsize{$5$ iterations}}}
    \put(23,74.2){\makebox(0,0)[t]{\scriptsize{$20$ iterations}}}
    \put(40,55){\makebox(0,0)[t]{\scriptsize{PIC}}}
    \put(59.6,42.5){\makebox(0,0)[t]{\scriptsize{MMSE}}}
    \put(82,58){\makebox(0,0)[t]{\scriptsize{list det.}}}
    \put(81.3,55){\makebox(0,0)[t]{\scriptsize{with $T$-alg}}}
    \put(55,0){\makebox(0,0)[t]{number of users $K$}}
    \put(-1,45){\makebox(0,0)[r]{\rotatebox{90}{BER}}}
  \end{picture}}
  \caption{Comparative CDMA system BER performance as a function of 
    number of users $K$. Spreading gain $L=8$, $E_b/N_0=5$ dB.}
  \label{fig:loadplot}
\end{figure}

The performance of the $T$-algorithm in the above model is shown as a
function of the number of users in Figure \ref{fig:loadplot}, at
$E_b/N_0=5$ dB.  Also shown is the performance of two linear filters
commonly used as the multi-user detector in such systems, the PIC
\cite{AleGraRee98} and the LMMSE \cite{WanPoo99} filters.  The
parameter $\Pmin$ for the $T$-algorithm is set to $32$ for $K\le 16$,
$\Pmin=64$ for $K=17,18$, and $\Pmin=128$ for $K=19,20$ users.  These
values were found by experiment to be sufficiently large for the loads
considered.

While very computationally efficient, the PIC can only support $9$ users
after $20$ iterations, and is clearly not suitable for highly loaded systems.
The MMSE filter performs better, supporting $14$ users after $20$ 
iterations, but requires a matrix inversion per user per iteration.

Estimating the detector APP is a highly non-linear calculation, and 
the linearised models and
assumptions used by the above filters are not necessarily valid in the
model under consideration.  The performance of the $T$-algorithm in
Figure \ref{fig:loadplot} clearly demonstrates the performance
advantage of approximating the APP directly using the rules of
probability.  List-detection using the $T$-algorithm supports $16$
users with only $5$ iterations and $19$ users after $20$ iterations at
$5$ dB.  To our knowledge we have not seen loads in such a system
approaching those achieved here.  

\begin{figure}[htbp]
  \centering\small\setlength{\unitlength}{.72mm}
  \scalebox{.975}{
  \begin{picture}(100,80)
    \put(0,3){\includegraphics[width=72mm]{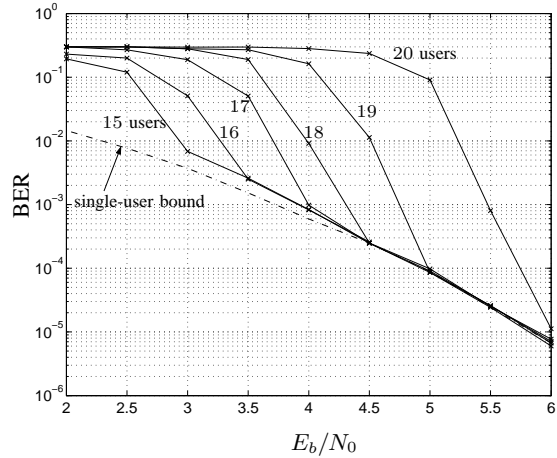}}
    \put(20,45){\makebox(0,0)[t]{\scriptsize{single-user bound}}}
    \put(19,60){\makebox(0,0)[t]{\scriptsize{$15$ users}}}
    \put(37,58){\makebox(0,0)[t]{\scriptsize{$16$}}}
    \put(39,63){\makebox(0,0)[t]{\scriptsize{$17$}}}
    \put(53,58){\makebox(0,0)[t]{\scriptsize{$18$}}}
    \put(63,62){\makebox(0,0)[t]{\scriptsize{$19$}}}
    \put(74,73){\makebox(0,0)[t]{\scriptsize{$20$ users}}}
    \put(55,0){\makebox(0,0)[t]{$E_b/N_0$}}
    \put(-1,45){\makebox(0,0)[r]{\rotatebox{90}{BER}}}
  \end{picture}}
  \caption{CDMA system BER performance after $20$ iterations as a function 
    of SNR. 
    Spreading gain $L=8$.}
  \label{fig:perf20}
\end{figure}

In Figure \ref{fig:perf20} is shown the performance of list 
detection as a function of $E_b/N_0$, for various number of users, 
after $20$ receiver iterations.
Note that without the log-likelihood transformation of Section \ref{sec:app},
the tree search with $20$ users would require at least 
one stage with $2^{K-L}=4096$ node computations, assuming the best case
that $\S$ has rank $L$. An extrinsic-information transfer chart \cite{exit} 
shows that the 
$T$-algorithm detector is very well matched in shape to the particular code, 
which helps to explain the good performance after many iterations, 
even at very high loads.
The charts, which we do not include here, also predict very accurately the 
convergence characteristics shown in Figures \ref{fig:loadplot} and
\ref{fig:perf20}.


\begin{figure}[htbp]
  \centering\small\setlength{\unitlength}{.72mm}
  \scalebox{.975}{
  \begin{picture}(100,83)
    \put(-11,-5){\includegraphics[width=90mm]
      {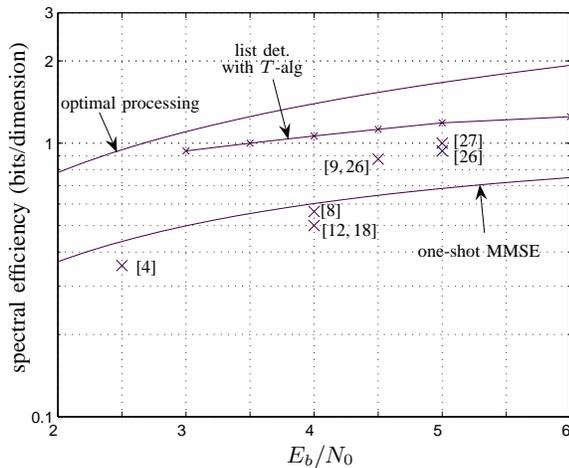}}
    \put(85,38){\makebox(0,0)[t]{\scriptsize{one-shot MMSE}}}
    \put(19,66){\makebox(0,0)[t]{\scriptsize{optimal processing}}}
    \put(44,76){\makebox(0,0)[t]{\scriptsize{list det.}}}
    \put(44,73){\makebox(0,0)[t]{\scriptsize{with $T$-alg}}}
    \put(22,35){\makebox(0,0)[t]{\scriptsize{\cite{ValWoe98PIMRC}}}}
    \put(57,45.8){\makebox(0,0)[t]{\scriptsize{\cite{AleGraRee98}}}}
    \put(60.5,42){\makebox(0,0)[t]{\scriptsize{\cite{ReiGraAle02,KuhHag04}}}}
    \put(60,54){\makebox(0,0)[t]{\scriptsize{\cite{WanPoo99,RasGraAle04}}}}
    \put(83,56){\makebox(0,0)[t]{\scriptsize{\cite{RasGraAle04}}}}
    \put(83,59){\makebox(0,0)[t]{\scriptsize{\cite{TanRas04}}}}
    \put(55,0){\makebox(0,0)[t]{$E_b/N_0$}}
    \put(0,45){\makebox(0,0)[r]{\rotatebox{90}{spectral efficiency 
          (bits/dimension)}}}   
  \end{picture}}
  \caption{Maximum spectral efficiency achieved by the optimal receiver, 
    one-shot LMMSE filter, and the iterative list-detection receiver for the 
    system described above, as a function of $E_b/N_0$.
    Code rate $R=1/2$.
    Also shown as crosses are rates achieved in the randomly spread 
    CDMA channel by receivers described in the literature.
    The primary work describing each receiver is referenced directly
    on the plot.}
    \label{fig:spectral_efficiency}
\end{figure}

Figure \ref{fig:spectral_efficiency} shows the spectral efficiency of
the receiver for the system described above, as a function of
$E_b/N_0$, measured as the maximum number of users for which
single-user performance is reached.  Also shown is the maximum
spectral efficiency $C$ achievable by using both an optimal joint
receiver, and an MMSE detector followed by single-user decoding.
These curves were approximated using the large-systems expressions for
random spreading given in \cite{VerSha99} under the constraint
$C=KR/L$ with $R=1/2$.

The receiver easily approaches optimal joint-processing data rates at
low SNR, but cannot maintain this slope with increasing SNR.
Nevertheless, the iterative $T$-algorithm receiver outperforms any
other practical algorithm we are aware of in terms of
system load for the given channel model.
Various other results for randomly spread CDMA in AWGN using a rate 1/2 code
are available in the literature, utilising the same canonical 
receiver structure but differing in the multi-user detector implementation.
These are shown in Figure \ref{fig:spectral_efficiency}, along
with references to the relevant papers.

\section{Conclusion}\label{sec:conclusion}
We have shown that near-optimal performance may be achieved with low
complexity in a randomly spread CDMA channel by employing the turbo
principle in an iterative receiver.  This is not a new observation;
our contribution is to show that by attempting to calculate the true
symbol-APP distributions in the inner detector, the performance is
significantly improved over detectors that employ linear filters, or
other structures derived using alternative considerations.  Close
approximation of the desired APP distributions in overloaded or
singular channels is practically facilitated by the simple procedure
of Section \ref{sec:app}, and by the application of simple and
well-known sequential algorithms.

\bibliographystyle{IEEE}
\bibliography{reference}

\end{document}